\documentclass[12pt]{iopart}
\usepackage{graphicx}
\usepackage{bm}
\usepackage{epsfig,subfigure}

\newcommand{\AuAu}{{$Au$-$Au$\ }}
\newcommand{\CuCu}{{$Cu$-$Cu$\ }}
\newcommand{\PbPb}{{$Pb$-$Pb$\ }}

\newcommand{\GeVc}{${\rm GeV}/c$\ }

\newcommand{\ds}{{\displaystyle}}

\begin{document}

\title{Parton-medium cross section and average QGP viscosity 
 in lead-lead collisions at LHC}



\author{V.~L.~Korotkikh$^1$, E.~E. Zabrodin$^1$, $^2$}

\vskip 10mm

\address{$^1$Skobeltzyn Institute for Nuclear Physics, Moscow State 
University, Russia }
\address{$^2$Department of Physics, University of Oslo, Norway}

\eads{\mailto{vlk@cern.ch}}

\date{Received: date / Accepted: date}
\begin{abstract}
The recent experimental CMS and E-by-E ATLAS data on the elliptic 
anisotropy of particles in \PbPb collisions are used to extract parton 
medium cross-section and the quark-gluon plasma viscosity-to-entropy 
ratio within the incomplete equilibration medium model. 
Our method assumes extrapolation of the measured pseudorapidity 
spectrum of charged particles to low-$p_T$ range. Then the rapidity
distribution of both charged and neutral hadrons is restored. The
extracted value of the parton-medium cross section is found to be 
$\ds \sigma = (3.1 \pm 0.2) {\rm mb}$. It yields $\eta/s = 0.17 \pm 
0.02$ for the ratio of shear viscosity to entropy density in QGP
phase. The last result is in a good agreement with the estimates 
obtained by calculations within the viscous hydrodynamics.
\end{abstract}



\maketitle

\section{Introduction}
\label{sec:intro}

Observations at the RHIC and LHC suggest that an extremely dense 
partonic medium, most probably quark-gluon plasma (QGP), with nearly
perfect fluid properties has been formed at the extreme temperatures and 
energy densities produced in relativistic heavy ion collisions. 
As the experimental precision was instantly improving, it became obvious 
soon that the description of the data on relativistic heavy ion 
collisions required some degree of QGP viscosity 
(for review see, e.g., \cite{Heinz_Snell}).

Theoretical works on strongly coupled quantum field theories, which 
utilize results from superstring theory, have established a lower limit 
of about $1/4\pi$ for the specific shear viscosity $\eta/s$ of the QGP 
\cite{Policastro}. Here $\eta$ is the shear viscosity and $s$ is the 
entropy density in the system. QGP shear viscosity is an important 
parameter characterizing the QGP medium.


One of the possible ways to extract the QGP viscosity is to confront the 
data with the macroscopic model calculations. Here the shear viscosity,
or rather $\eta/s$ ratio, is inserted as a free parameter in the 
hydrodynamic part of the code. Its value is fixed after providing the
best agreement with the data. The most successful approaches, 
however, combine a viscous fluid dynamic description of the QGP phase 
with a microscopic Boltzmann simulation of the hadronic phase 
\cite{Song_2011,Ryu_2012}. For instance,  
VISHNU model \cite{Song_2011} matches the (2+1)-di\-men\-sio\-nal, 
longitudinally boost-invariant viscous hydrodynamic algorithm VISH2+1 
\cite{Shen:2011kn,Song_2008} to the well-known UrQMD cascade model 
\cite{Bass_1998} as an afterburner. In a similar approach, MUSIC+UrQMD 
\cite{Ryu_2012}, the evolution is fully (3+1)-di\-men\-sio\-nal. 
Calculations within this model, performed with the IP-Glasma 
initial conditions \cite{Schenke_2012} together with an average value 
$\eta/s = 0.2$ for \PbPb collisions at the LHC and a somewhat smaller 
value $\eta/s = 0.12$ for \AuAu collisions at top RHIC energies, 
provide a good description of all presently available data for charged 
hadron flow harmonics, both $v_n$ and $v_n(p_T)$ from $n = 2$ to 
$n = 5$ \cite{Gale_2012}, including their centrality and $p_T$- 
dependencies. Usually this approach is called the viscous hydrodynamic 
hybrid model.

Other methods link the $\eta/s$ ratio to the system parameters, such
as temperature and partonic cross section, which can be obtained from
the analysis of particle spectra.
The eccentricity scaled integrated elliptic flow $v_2/\varepsilon$ 
versus transverse charged-particle density $n_T=(1/S)(dN_{ch}/dy)$ is 
a ``universal'' function, since this function describes energy and 
centrality dependencies of the flow in a wide range of collision 
energies. It is also less sensitive to the initial conditions of heavy 
ion collisions. There are many examples of application of this 
``universal'' function for data analysis, see, e.g.,
\cite{Song_2011,Shen:2011kn,Bhalerao_1,Drescher:2007cd,Chaudhuri,Liu}.
One can also note a good agreement with the numerous experimental data 
on $v_2/\varepsilon$ as a function of $1/S~dN/d\eta$ from 7.7\,AGeV to 
2.76\,ATeV for the Monte Carlo Glauber (MC-Gl) initial conditions in 
contrast to the Kharzeev-Levin-Nardi (MC-KLN) ones ~\cite{Shen:2011kn}.
In \cite{Bhalerao_1} the authors suggest a simple equation for the 
``universal'' function as a function of transverse particle density 
within the incomplete equilibration medium (IEM) approach. This equation 
was directly used in \cite{Drescher:2007cd,Chaudhuri} to extract the IEM 
parameters from the RHIC data at 200 AGeV. To our best knowledge, no 
similar analysis of the LHC data has been done yet.


 Hadrons, as well as partons, produced in relativistic heavy ion 
collisions interact further with their neighbors, thus developing a
collective flow which has both isotropic and anisotropic components.
The strength of the collective anisotropic flow is measured 
by means of harmonics of the Fourier expansion of the charged hadron 
azimuthal distributions with respect to the event plane, defined by the 
maximum particles in direction of the harmonic studied.
The azimuthal dependence of the particle yield can be written
in terms of the harmonic expansion \cite{Poskanzer:1998yz}
\begin{eqnarray}
\label{Intr_eq:1}
\ds
& & E\frac{d^3N}{dp^3}=\frac{d^2N}{2\pi p_{\rm T}dp_{\rm T}d\eta} 
\times \nonumber \\
& & \left\{ 1 + 2\sum\limits_{n = 1}^\infty v_{\rm n}(p_{\rm T},\eta)  
\cos{[n(\varphi -\Psi_{\rm n})]} \right\}~,
\end{eqnarray}
where $\varphi$ is the azimuthal angle with respect to the event plane 
$\Psi_{\rm n}$, and $v_n$ is the  Fourier coefficient. In this paper a 
parameter $v_2$, dubbed elliptic flow, is considered. Elliptic flow was 
studied extensively at both LHC \cite{ALICE_v2,ATLAS_EbyE,CMS_v2}
and RHIC 
\cite{Adler:2003kt,Adams:2003am,Afanasiev:2007tv,Abelev:2007qg} 
energies. Its description is amenable to hydrodynamic calculations
\cite{Teaney:review,Kolb:2003dz,Huovinen:2006jp,Heinz:2009xj}.
Comparisons to theoretical results indicate that at the LHC energies 
most of the flow is primarily generated during an early stage of the 
system evolution before the hadronization, and the relative impact of 
the late hadronic stage is weak. An integrated elliptic flow can be 
expressed as
\begin{equation}
\label{Intr_eq:2}
\ds
v_2 = \frac{\int_{\Delta y} dy \int_{\Delta p_T} p_T dp_T 
\frac{{d^2 N}}{{p_T dp_T dy}}(p_T,y)v_2(p_T,y)}
{\int_{\Delta y} dy \int_{\Delta p_T} p_T dp_T 
\frac{{d^2 N}}{{p_T dp_T dy}}(p_T,y)}.  
\end{equation}
Several methods have been developed to extract the elliptic flow from 
the experimental data. In present paper we utilize the approach 
described in the CMS publication \cite{CMS_v2}.

Recent theoretical studies of elliptic flow were also focused on the 
quantifying of the ratio of the shear viscosity to the entropy density 
of the produced medium in the framework of viscous hydrodynamics
\cite{Teaney:review,Romatschke:2009im} with a variety of possible 
initial conditions. 
In the theory of strongly interacting systems 
\cite{Teaney:review,Plumari_2012} for the case of isotropic cross 
section and massless particles the QGP shear viscosity $\eta$ is related 
to the scattering cross-section $\sigma$ via the equation 
\begin{equation}
\label{visc}
\ds
\eta = 1.267 ~ \frac{T}{\sigma}
\end{equation}
in natural units system, where $\hbar  = 1$ and $c = 1$. 
There are also higher order calculations extended up to the 16th order 
that do not change the obtained result \cite{Wiranata_2012}.
For the gluonic QGP which obeys the Boltzmann distribution, the entropy 
density is given by $s=4 n$, where $\ds n=\frac{g}{\pi^2}T^3$ is the QGP 
medium density and $g = 16$ is the gluon degeneracy. Thus, the ratio of 
shear viscosity to entropy density may be written as
 
\begin{equation}
\label{eq:1a}
\ds
\frac{\eta}{s} = 0.194 ~ \frac{1}{T^2~\sigma}~.
\end{equation}

We assume that the anisotropic flow is formed until the freeze-out of 
the fireball and it is not changed further by the final-state 
interactions of hadrons. Therefore, we need to know $T = T_{freez}$, 
which can be obtained from the experimental data. We used the standard 
Statistical Hadronization Model (SHM) with the modification factors 
from the UrQMD \cite{Andronic_2012} to fit the yields of identified 
charged hadrons produced in \PbPb collisions at $\sqrt{s_{NN}}$ = 
2.76\,TeV. Data are taken from the ALICE experiment at the LHC 
\cite{ALICE_v2}. The resulting value of the temperature of chemical 
freeze-out, $T_{freez} = 166 \pm 3$\,MeV, is compatible with that at 
RHIC energies and expectations from the lattice QCD calculations 
\cite{Becattini_2012,Stock_2013}.

The paper is organized as follows. The features of analysis of 
the azimuthal harmonics on the Event-by-Event (E-by-E) basis are 
sketched in Sec.~\ref{sec:EbyE}. The model of incomplete equilibration 
medium (IEM) is considered in Sec.~\ref{sec:model}. 
Section~\ref{sec:ellipflow} presents details of our analysis of 
experimental data and main results obtained for parton-medium cross 
section and share viscosity-to-entropy ratio. Conclusions are drawn in 
Sec.~\ref{sec:summ}.

\section{Event-by-Event analysis}
\label{sec:EbyE}

The Event-by-Event analysis was implemented by ATLAS Collaboration to
study the azimuthal harmonics in \PbPb collisions at $\sqrt s_{NN}$ = 
2.76 TeV, see \cite{ATLAS_EbyE,Jia_1,Jia_2,Jia_3}. 
Here the density distribution $P(v_n)$ was unfolded by the standard 
Bayesian procedure. It was shown that the true parameters density 
distribution  did not depend on experimental 
extraction methods such as event plane, cumulant 
or two particle correlation with the triggering particle. 
The E-by-E analysis also excludes nonflow contributions.

If $v_2$ is understood as the hydrodynamic response to the initial
anisotropy, $\varepsilon$, of the initial density profile, then 
$v_n=k_n \times \varepsilon_n$. In \cite{Jia_1,Jia_2,Jia_3} the 
authors suggested a simple functional Elliptic Power form of 
the initial state  without assuming any particular model of the initial 
conditions, which fitted eccentricity density distribution better than 
Gaussian or Bessel-Gaussian distribution.

The Elliptic Power distribution of eccentricity reads \cite{Jia_3}
\begin{equation}
\label{dist_ecc}
\ds
p(\varepsilon_n) =  \frac{2\alpha\varepsilon_n}{\pi}(1-
\varepsilon_0^2)^{\alpha+0.5}\int_0^\pi
\frac{ (1-\varepsilon_n^2)^{\alpha-1} d\varphi}{(1-\varepsilon_0
\varepsilon_n ~cos\varphi)^{2\alpha+1}}
\end{equation}
The probability distribution of anisotropic flow, $P(v_n)$, is related 
to the distribution of the initial anisotropy $p(\varepsilon_n)$ via
\begin{equation}
\label{dist_flow}
\ds
P(v_n) =  \frac{1}{k_n} p\left( \frac{\varepsilon_n}{k_n} \right) \ ,
\end{equation}
where $k_n$ is a response parameter. Below we will use the notation 
$k_n=k_n^{EbyE}$.

\section{Incomplete equilibration medium model }
\label{sec:model}

Deviation from the local equilibrium can lead to an indicative 
dependence of the eccentricity scaled elliptic flow on charged particle 
multiplicity. Local equilibrium is not a necessary condition for the 
elliptic flow, but it is commonly accepted that deviation from the
equilibrium can only reduce the magnitude of the effect.

Degree of thermalization in the fluid, produced in heavy ion collisions, 
can be characterized by the dimensionless parameter, Knudsen number $K$ 
\cite{Knudsen}. The Knudsen number $K=\lambda/R_{tr}$ is defined by 
evaluating the mean free path $\lambda$ at average QGP life time 
$\tau = R_{tr}/c_s$, with the transverse medium size, $R_{tr}$, and the 
speed of sound in ultrarelativistic medium, $c_s = 1/\sqrt{3}$.
By definition, $K^{-1}$ is the mean number of collisions per particle 
(also known as ``opacity''); thus the ideal hydrodynamic limit 
corresponds to $K \rightarrow 0$. Hence, $\lambda=1/(\sigma\,\rho)$, 
where $\sigma$ is the effective parton-medium (mostly gluon-gluon) cross 
section and $\rho$ is the (time-dependent) density of the medium for a 
typical time $\tau$. 

The inverse Knudsen number  $K^{-1}$ can be determined from the 
experimental data \cite{Bhalerao_1} . It is proportional to transverse 
particle density at midrapidity
$\ds n_T=\frac{1}{S}\frac{dN}{dy}\Big{|}_{y=0}$, where 
$\ds \frac{dN}{dy}\Big{|}_{y=0}$ 
is the  multiplicity density, an indicator of the number of collisions 
per particle at the time when elliptic flow is formed, and $S$ is a 
square of the transverse area of the collision zone. Therefore, the 
inverse Knudsen number reads
\begin{equation}
\label{eq:1}
\ds
K^{-1}(n_T)~=~c_s\sigma~n_T~=~c_s\sigma~\frac{1}{S}\frac{dN}{dy} 
\Big{|}_{y=0}\ .
\end{equation}

In the incomplete equilibration medium (IEM) model the measured ratio 
$v_2/\varepsilon$ is related to the Knudsen number by a simple equation 
suggested in \cite{Bhalerao_1,Borg}
\begin{eqnarray}
\ds
\label{eq:2}
\frac{v_2}{\varepsilon}(n_T)~&=&~\Bigl{(}\frac{v_2}{\varepsilon} 
\Bigr{)}^{\rm max}~\frac{K^{-1}}{K^{-1}+K_0^{-1}} \nonumber \\
 ~&=&~ \Bigl{(} \frac{v_2}{\varepsilon}\Bigr{)}^{\rm max}~
\frac{1}{1+K/K_0}\nonumber \\ 
 ~&=&~ \Bigl{(}\frac{v_2}{\varepsilon}\Bigr{)}^{\rm max}~
\frac{n_T}{n_T+1/(c_s\sigma K_0)}\ ,
\end{eqnarray}
where $\ds \left( \frac{v_2}{\varepsilon} \right)^{\rm max}$ represents 
the hydrodynamic limit. Local thermal equilibrium is achieved if 
$K^{-1} \gg 1$. The value of parameter $K_0 = 0.70 \pm 0.03$ is 
estimated from the Monte-Carlo simulation of transport equations 
\cite{Gombeaud}.
As seen in Eq.(\ref{eq:2}), using the known values of $c_s$ and $K_0$,  
we can extract from the experimental data two parameters 
$\Bigl{(}\frac{v_2}{\varepsilon}\Bigr{)}^{\rm max}$ and $\sigma$, or, 
alternatively, the combined parameter $(c_s\sigma K_0)$. 
The success of IEM model, which gives the universal function, is caused 
perhaps by a fact that the physical uncertainties of the
individual parameters $c_s$, $\sigma$, and $K_0$ are canceled out in 
their product, $c_s\sigma K_0$.

Elliptic flow develops gradually in the system as the system evolves. 
Measured value of $v_2$ is formed at the final stage of hydro evolution 
at the temperature of freeze-out. The quantities that we shall extract 
from IEM model should be interpreted as the averages over the transverse 
area S, and over some time interval around $\tau = R_{tr}/c_s$, which is 
the typical time scale for the build-up of $v_2$ in hydrodynamics.

There are several works \cite{Drescher:2007cd,Chaudhuri,Liu}, where 
Eq.~(\ref{eq:2}) is applied for the parameter extraction.
First analysis of the experimental data for \AuAu and \CuCu collisions 
at $\sqrt{s_{NN}} = 200$\,GeV within the IEM model was made in 
\cite{Drescher:2007cd}.
The authors determined parameters $v_2^{\rm max}$ and $\sigma$ 
for different initial conditions, such as MC-Gl and MC-KLN ones.
Later on, in \cite{Chaudhuri} the Eq.~(\ref{eq:1}) was generalized by 
including the ratio of the shear viscosity to entropy density:
\begin{equation}
\label{eq:3}
\ds
K^{-1}(n_T)~=~c_s\sigma~ n_T 
~\Bigl{[}1+\frac{2}{3\tau_iT_i} ~\left(\frac{\eta}{s} \right)~
\Bigr{]}^{-3}\ .
\end{equation}
However, the parameters $\eta/s$ and $\sigma$ are strongly correlated, 
as can be easily seen from the relation
$$\ds \sigma~=~const \cdot  ~\Bigl{[}1+\frac{2}{3\tau_iT_i} ~
\left(\frac{\eta}{s} \right)~\Bigr{]}^{3}.$$
It means that any fit will provide us the same $\chi^2/NDF$ for two 
different pairs of parameters, $\eta/s$ and $\sigma$. Therefore, this 
model is not used in our analysis.

\section{Eccentricity-scaled elliptic flow as a function of transverse 
particle density}
\label{sec:ellipflow}

For the analysis we use the experimental data on the azimuthal 
anisotropy in \PbPb collisions at $\sqrt{s_{NN}}= 2.76$\,TeV  obtained  
with the Compact Muon Solenoid (CMS) detector at the LHC \cite{CMS_v2}. 
The final data sample (with all the necessary cuts applied) contains 
22.6 million minimum-bias events, corresponding to the  integrated 
luminosity of approximately 3 $\mu b^{-1}$.
 
The $v_2$ coefficient is determined as a function of charged particle 
transverse momentum $p_T$ and overlap of the colliding nuclei (centrality) 
in the pseudorapidity region of $|\eta|~<~2.4$, where $\eta = 
-\ln(\tan(\theta/2))$, with $\theta$ being the polar angle relative to 
the counterclockwise beam direction. We analyze the integrated elliptic 
flow in the measured transverse momentum range of $0.3 \leq$ $p_T$ 
$\leq 3.0$\,\GeVc and extrapolate it to the region 
$0 \leq$ $p_T$ $\leq 0.3$\,\GeVc.

The collective motion of the system and, therefore, the anisotropy 
parameter depends on the initial shape of the nucleus-nucleus collision 
area and the fluctuations in the positions of the interacting nucleons. 
By dividing $v_2$ to the participant eccentricity, one may potentially 
remove this dependence across centralities, colliding species, and 
center-of-mass energies, thus enabling a comparison of results in terms 
of the underlying physics driving the flow.
Two initial state definitions are used in MC-Gl and MC-KLN models. 
Initial state in the MC-KLN model 
\cite{Hirano:2005xf,Hirano:2010jg,Song:2010mg,Nagle:2011uz,Shen:2011eg,Qiu:2011iv} 
is based on the color glass condensate (CGC) concept 
\cite{PhysRevC.76.041903} accounting for the fact that at very high 
energies or small values of Bjorken $x$ the gluon density becomes 
saturated at very high values.


The CGC model predicts eccentricities that exceed the Glauber-model ones 
by a factor of about 1.2, with some deviations from this value in the 
most central and the most peripheral collisions
\cite{Nagle:2011uz,Qiu:2011iv}, and a bit different behavior of the 
initial overlap area $S$ ~\cite{Shen:2011kn}. 

In \cite{Shen:2011kn} it was shown that only in MC-Gl model the
excellent universal curve propagates over the entire collision energy 
range between 7.7~AGeV and 2.76~ATeV for the eccentricity scale 
depending on $1/S~dN_{ch}/d\eta$. Therefore, we use the Glauber model, 
which is a multiple-collision model treating a nucleus-nucleus 
collision as an independent sequence of nucleon-nucleon collisions (see 
\cite{Miller} and references therein), to calculate the participant 
eccentricity 
$$\varepsilon_\mathrm{part}=\frac{ \sqrt { \left( \sigma_{y'}^{2} - 
\sigma_{x'}^{2} \right)^{2} + 4\sigma_{x'y'}^{2} } }{ \sigma_{y'}^{2} + 
\sigma_{x'}^{2} }$$ 
and its cumulant moments 
$\ds \varepsilon\{2\}=\sqrt{\langle \varepsilon_\mathrm{part}^{2} \rangle}$, 
where $\sigma_{y'}^{2}$ and $\sigma_{x'}^{2}$ are the variances of the 
participant spatial distribution and $\sigma_{x'y'}^{2}$ is the 
covariance. The participant eccentricity is evaluated in a frame that 
maximizes its value. This frame, which defines the initial-state 
participant plane, may be shifted and rotated with respect to the frame 
defined by the impact parameter vector and the beam direction. From the 
Glauber model, we also calculate the transverse overlap area of the two 
nuclei 
$S \equiv \pi \sqrt{\sigma_{x'}^{2}\sigma_{y'}^{2}-\sigma_{x'y'}^{2}}$.
The table of eccentricities $\varepsilon$ and overlap area $S$ for each 
centrality can be found in ~\cite{CMS_v2}. 


To restore the elliptic flow the Event Plane (EP) method is employed. 
Here the two-particle positive and negative event planes, $\Psi_2^+(EP)$ 
and $\Psi_2^-(EP)$, are determined by correlating the signals in the 
hadron calorimeter HF+ $(3 < \eta < 5)$ and HF$-$ $(-5 < \eta < -3)$, 
respectively. The elliptical anisotropies are measured by correlating 
particles from the negative (positive) $\eta$ range of the tracker, 
i.e., $-2.4 < \eta < 0 ~(0 <\eta < 2.4)$ to $\Psi_2^+(EP)$ 
$(\Psi_2^-(EP))$. The factor $R$ corresponds to a resolution correction, 
which needs to be applied to the event plane. A gap of three units in 
$\eta$ is kept to reduce the contamination of non-flow effects.

In \cite{CMS_v2} the isotropic parameter $v_2$ was measured at $|\eta| <
0.8$ in the region $p_T = 0.3 - 3.0$\,\GeVc. Now we extrapolate the 
integrated value of $v_2$ for every centrality to the region $p_T = 0 - 
0.3$\,\GeVc, using the extrapolation of the normalized spectrum 
$dN/dp_T$ at each 
centrality by power-law dependence and polynomial of 5-th order fit to 
$v_2(p_T)$ in the region $p_T = 0.3- 3.0$\,\GeVc. 
 Such procedure allows us to calculate the 
integral value of $v_2$ by using Eq.~(\ref{Intr_eq:2}) in the region 
$0 < p_T < 3$\,\GeVc.


The eccentricity scaled integrated elliptic flow, $v_2/\varepsilon$,
extracted by the two-particle cumulant method is nearly identical to the 
flow obtained by the event-plane method, provided the corresponding 
eccentricity fluctuations are taken into account.
In the centrality range of 15 -- 40\%, the four-particle cumulant 
measurement of $v_2\{4\}/\{\varepsilon\{4\}$ is also in agreement with 
the other two methods. The main difference in the results obtained by 
different methods can be attributed to their sensitivity to nonflow 
contributions. Because of some irregularities of eccentricity 
$\varepsilon\{4\}$ for very central and for peripheral collisions, we do 
not include the four-particle cumulant in the analysis of transverse 
particle density in present work.

We have also recalculated $dN/dy$ from $dN/d\eta$ at $|\eta| < 0.8$.
Under the assumption of region mid-rapidity, i.e., $cosh^2 y \approx 
1.0$, the charged particle spectrum $dN/dp_T$ is used for each 
centrality, and $dN/dy$ is calculated as
\begin{eqnarray}
\ds
\frac{dN}{dy} &=& \int dp_T\frac{1}{1-\frac{m^2}{m^2_Tcosh^2y}}
\frac{dN}{d\eta dp_T}
\nonumber \\
&\approx& \int dp_T\frac{m_T}{p_T} \left( \frac{dN}{dp_T}\Big{|}_{\eta=0} 
\right)\ ,
\end{eqnarray}
where $m_T^2=m+p_T^2$. Recall, that the analysis in \cite{CMS_v2} is 
done for charged particles, where the spectrum is dominated by 
$\pi^\pm$'s. In present work we make two modifications. Namely, (i) we 
put $m=m_{\pi^0}$ and (ii) recalculate the total number of particles as 
$dN/dy=3/2dN_{ch}/dy$, thus taking into account neutral $\pi$-mesons.



In ideal hydrodynamics, the eccentricity-scaled elliptic flow is
constant over a broad range of impact parameters. The deviations from 
this behavior are expected in peripheral collisions, in which the system 
freezes out before the elliptic flow fully builds up and saturates 
\cite{Kolb:2000sd}. A weak centrality and beam-energy dependence is
expected through variations in the equation-of-state. In addition, the
system is also affected by viscosity, both in the sQGP and in hadronic 
stages~\cite{Shen:2011kn,Heinz:2009xj,Hirano:2010jg,Luzum:2009sb} of its 
evolution. Therefore, the centrality dependence of $v_{2}/\varepsilon$ 
can be used to extract the ratio of the shear viscosity to the entropy 
density of the system.

It was previously observed \cite{PHOBOSeccPART,PhysRevC.68.034903} that 
the $v_2/\varepsilon$ values, obtained for different systems colliding at
varying beam energies, scale with the charged-particle rapidity density 
per unit transverse overlap area, $(1/S) (dN_{\mathrm{ch}}/dy)$, which 
is proportional to the initial entropy density. In addition, it was 
pointed out \cite{Song:2010mg} that in this representation the 
sensitivity to the modeling of the initial conditions of heavy ion 
collisions is largely removed, thus enabling the extraction of the 
shear viscosity to the entropy density ratio from the data via the 
comparison with viscous hydrodynamic calculations.  



\begin{figure*}

\begin{center}

\resizebox{0.8\textwidth}{!}{%

\includegraphics{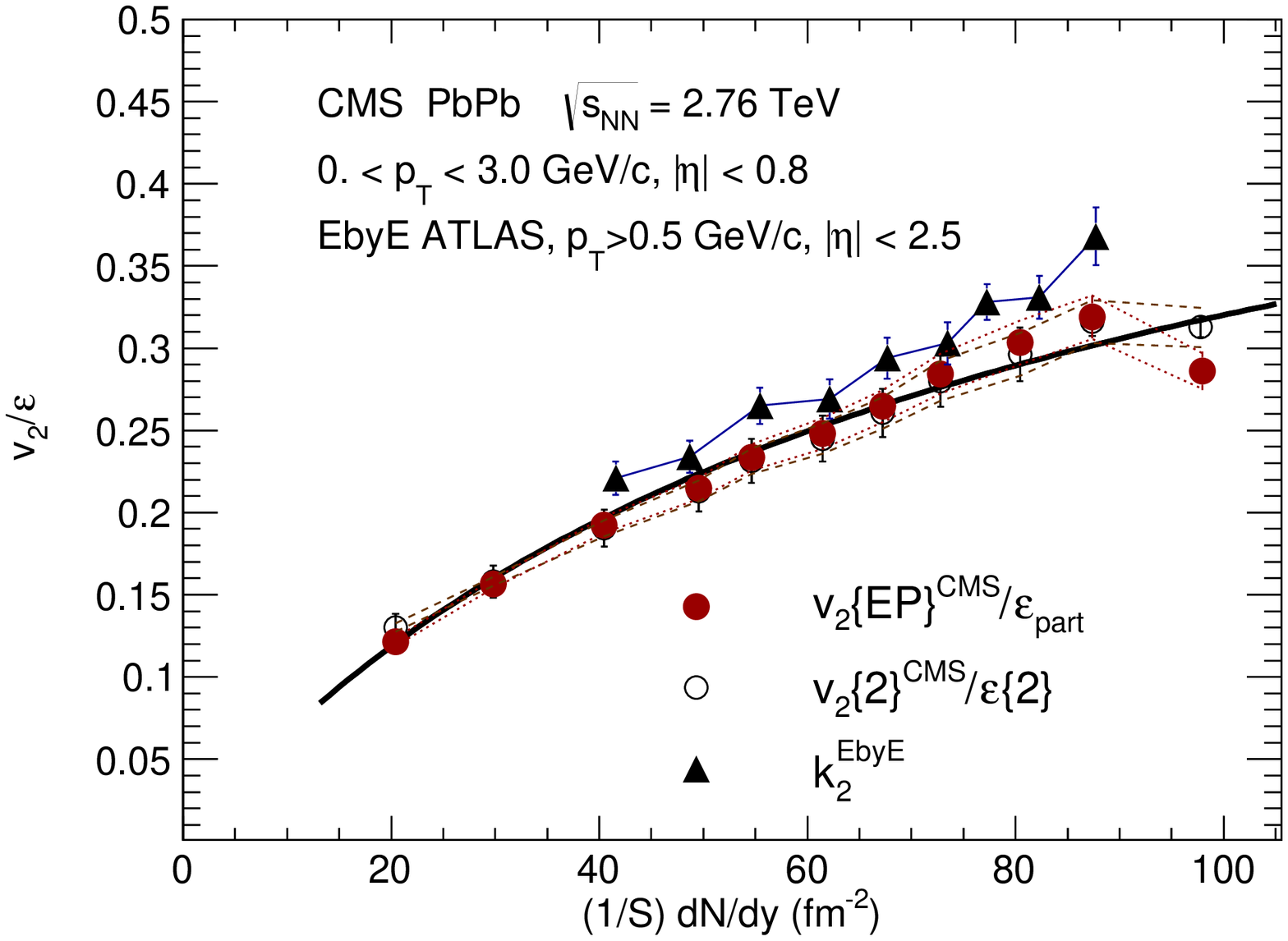}

}

\end{center}

\caption{Color online.
Eccentricity-scaled $v_{2}\{EP\}/\varepsilon_{part}$ (full circles) and  
$v_{2}\{EP\}/\varepsilon{\{2\}}$ (open circles) from CMS 
\protect\cite{CMS_v2}  as a function of total particle transverse density 
$1/S d N/d y$ extended to $p_T$=0 region (see text for details). 
Full triangles denote the E-by-E results of ATLAS 
\protect\cite{ATLAS_EbyE} with extraction response parameter $k_2^{EbyE}$ 
from \protect\cite{Yan_14}. 
The full black line is a fit  of  $v_{2}\{EP\}/\varepsilon_{part}$   
to Eq.~(\protect\ref{eq:2}). The dashed lines represent the band of 
systematic uncertainties in the eccentricity determination.} 

\label{vlk}

\end{figure*}

In Fig.~\ref{vlk} we plot the CMS data on $v_2\{EP\}/\varepsilon_{part}$ 
and $v_2\{2\}/\varepsilon\{2\}$ measured in \PbPb collisions at
$\sqrt{s_{NN}} = 2.76$\,TeV \cite{CMS_v2}. Here the pseudorapidity 
spectrum of char\-ged particles, $dN_{ch}/d\eta$, was (i) extrapolated 
to the region $0 < p_T < 3$\,\GeVc, then (ii) converted to the rapidity 
distribution $dN_{ch}/dy$, which finally was used (iii) to obtain the 
total spectrum, $dN/dy=3/2~dN_{ch}/dy$, of charged and neutral hadrons.
Solid curve in Fig.~\ref{vlk} corresponds to the fit of
$v_2\{EP\}/\varepsilon_{part}$ distribution to Eq.~(\ref{eq:2}) with 
two free parameters, $\ds (\frac{v_2}{\varepsilon})^{max}$ and $\sigma$. 

The fitting procedure yields
\begin{equation}
\label{fit:2}
 ~\sigma~=~(3.1 ~\pm~ 0.2) {\rm mb}, \quad
\Bigl{(}\frac{v_2}{\varepsilon}\Bigr{)}^{max}=0.58~\pm0.03
\end{equation}
The indicated errors are only of statistical origin.
{\it
And for the viscosity-to-entropy ratio we get
\begin{equation}
\label{eq:5}
 \frac{\eta}{s} = 0.17 \pm 0.02 ~~~~~~ \textrm{or}  ~~~~~~ 
4\pi\Bigl(\frac{\eta}{s}\Bigr) = 2.14 \pm 0.25\ . 
\end{equation}

}

Let us discuss the obtained results and compare it with the available 
estimates. Since the lattice QCD calculations become inapplicable for
the non–equilibrium evolution, the common practice is  
to rely on the AdS/CFT correspondence for guidance to general properties 
of stron\-gly coupled field theories at finite temperatures 
\cite{Aharony}. AdS/CFT imposes a lower boundary on viscosity to entropy 
ratio as $4 \pi \eta/s \geq 1 $.
The upper limit of this ratio, estimated both from pure theoretical 
considerations and from the analysis of heavy ion collisions at 
energies of RHIC and LHC, is about 0.6 (for review, see 
\cite{TSchaef_14} and references therein).   
%
The hybrid hydrodynamic models provide $\eta/s \approx 0.16$ 
\cite{Song_2011} or even $\eta/s \approx 0.12$ \cite{Gale_2012} at
RHIC, and a bit higher values $\eta/s \approx 0.22 \pm 0.02$ 
\cite{Song_2011} and $\eta/s \approx 0.2$ \cite{Gale_2012} at LHC.

Recently, in \protect\cite{Yan_14} the authors applied 
Eq.~(\ref{dist_flow}) to ATLAS E-by-E results \protect\cite{ATLAS_EbyE} 
and got the eccentricity scaled elliptic flow $k_2^{EbyE} = v_2^{true}/
\varepsilon_2$ as a function of centrality. We plot in Fig.~\ref{vlk} 
their distribution for $k_2^{EbyE}$ (in ATLAS version with $k^\prime = 
0.10$), taken from Fig.~2 of \protect\cite{Yan_14}.
Compared to the CMS data, the ATLAS E-by-E points are located a bit 
higher because of absence of the $v_2(p_T)$ extrapolation to the 
low-$p_T$ region, $0 < p_T < 0.5$ GeV/$c$, which decreases the values
of the integrated flow. Nevertheless, two different methods of 
extraction the ratio $v_2/\varepsilon$ give rather close dependencies of
this ratio on transverse particle density. From the E-by-E analysis of
the ATLAS data the value $\eta/s \approx 0.19$ was estimated 
\cite{Yan_14}. It is remarkable that our result, obtained within the 
framework of incomplete equilibration model, agrees well with the 
hydrodynamic model estimates.

Other result is the combined parameter $c_s\sigma K_0$, which appears 
to be
\begin{equation}
\label{eq:6}
 c_s\sigma K_0 = (1.25 \pm 0.14)  \rm mb
\end{equation}
It might be also useful for pinning down the transport properties of QGP.

\section{Conclusions}
\label{sec:summ}
\noindent 

The incomplete equilibration medium model is used to extract the medium 
modified gluon-medium cross section and to calculate the ratio of shear
viscosity to entropy density. 
Both parameters represent the fundamental properties of QGP and play an
important role in relativistic kinetic theory and hydrodynamics.
Recent experimental CMS and ATLAS data on the 
elliptic anisotropy of charged particles in \PbPb collisions at 
$\sqrt s_{NN}$~=~2.76~TeV are employed for the analysis. The integrated 
elliptic flow is estimated by extrapolation of measured differential 
flow and momentum spectrum of charged hadrons in the range $0.3 < p_T < 
3$\,\GeVc to the range $0 < p_T < 3$\,\GeVc.
In contrast to earlier applications of the IEM model, here the rapidity 
distribution of charged particles was recalculated from their 
pseudorapidity spectrum. The total hadron multiplicity distribution is 
then taken as $dN/dy=3/2dN_{ch}/dy$ to account for both charged and 
neutral particles. The values of obtained parameters are 
$\sigma~=~(3.1 ~\pm~ 0.2)$\,mb for the partonic cross section and 
$\frac{\eta}{s} = 0.17 \pm 0.02$ for the ratio of shear 
viscosity to the entropy density, respectively.
The dependence of $v_2/\varepsilon$ on particle transverse density
obtained within the IEM model is very close to that obtained with 
the parameters from the E-by-E analysis of ATLAS data 
\cite{Yan_14}.
The extracted value of $\eta/s$ also agrees well with that found in
simulations of \PbPb collisions at LHC by the viscous hydrodynamic 
hybrid models, thus supporting the description of quark-gluon plasma
as a nearly perfect fluid.


\section*{Acknowledgments}
V.L.K. would like to express deep appreciation to the members of the CMS 
collaboration for fruitful cooperation. Especially, he would like to thank 
J.~Velkovska, S.~Sanders and S.~Tuo for active help and for providing the 
CMS data. We are grateful to our SINP colleagues A.M. Gribushin, 
I.P. Lokhtin, A.M. Snigirev, and D.N. Golovin for useful discussions. This 
work was supported in parts by the Grant of the President of Russian Federation for 
scientific Schools supporting (3042.2014.2).


\newpage


\section*{References}
\begin{thebibliography}{2}

\bibitem{Heinz_Snell} Heinz~U and Snellings~R 2013
{\it Annu. Rev. Nucl. Part. Sci.} {\bf 64} 123

\bibitem{Policastro} Policastro~G, Son~D~T and Starinets~A~O 2001 
{\it Phys. Rev. Lett.} {\bf 87} 081601

\bibitem{Song_2011} Song~H, Bass~S~A and Heinz~U 2011 
{\it Phys. Rev.} C {\bf 83} 024912

\bibitem{Ryu_2012} Ryu~S, Jeon~S, Gale~C, Schenke~B and Young~C 2013
{\it Nucl. Phys.} A {\bf 904-905} 389c

\bibitem{Shen:2011kn} Shen~C and Heinz~U 2011
{\it Phys. Rev.} C {\bf 83} 044909

\bibitem{Song_2008} Song~H and Heinz~U
2008 {\it Phys. Lett.} B {\bf 658}  279; \nonum 
2008 {\it Phys. Rev.} C {\bf 77} 064901 


\bibitem{Bass_1998}
Bass~S~A et al 1998  {\it Prog. Part. Nucl. Phys.} {\bf 41}  255);
\nonum
Bleicher~M et al 1999  {\it J. Phys. G: Nucl. Part. Phys.} {\bf 25} 1859

\bibitem{Schenke_2012} Schenke~B, Tribedy~P and Venugopalan~R 
 2012 {\it Phys. Rev. Lett.} {\bf 108} 252301; \nonum
 2012 {\it Phys. Rev.} C {\bf 86} 034908

\bibitem{Gale_2012} 
Gale~C, Jeon~S, Schenke~B, Tribedy~P and Venugopalan~R 2013 
{\it Phys. Rev. Lett.} {\bf 110} 012302

\bibitem{Bhalerao_1} 
Bhalerao~R~S, Blaizot~J~P, Borghini~N and Ollitrault~J-Y 2005 
{\it Phys. Lett.} B {\bf 627} 49

\bibitem{Drescher:2007cd}  
Drescher~H~J, Dumitru~A, Gombeaud~C and Ollitrault~J-Y 2007 
{\it Phys. Rev.} C {\bf 76} 024905

\bibitem{Chaudhuri} Chaudhuri~A~K  2010
{\it Phys. Rev.} C {\bf 82} 047901

\bibitem{Liu} Liu~J~L et al 2009 {\it Phys. Rev.} C {\bf 79} 064905

\bibitem{Poskanzer:1998yz} Poskanzer~A~M and Voloshin~S~A 1998 
{\it Phys. Rev.} C {\bf 58} 1671 
 
\bibitem{ALICE_v2} Aamodt~K et al (ALICE Collaboration) 2010
{\it Phys. Rev. Lett.} {\bf 105} 252302

\bibitem{ATLAS_EbyE} Aad~G et al (ATLAS Collaboration) 2012 
{\it Phys. Lett.} B {\bf 707} 330

\bibitem{CMS_v2} Chatrchyan~S et al (CMS Collaboration) 2013 
{\it Phys. Rev. } C {\bf 87} 014902 

\bibitem{Adler:2003kt} Adler~S~S et al (PHENIX Collaboration) 2003 
{\it Phys. Rev. Lett.} {\bf 91} 182301

\bibitem{Adams:2003am} Adams~J et al (STAR Collaboration) 2004 
{\it Phys. Rev. Lett.} {\bf 92} 052302

\bibitem{Afanasiev:2007tv} Afanasiev~S et al (PHENIX Collaboration) 2007 
{\it Phys. Rev. Lett.} {\bf 99} 052301

\bibitem{Abelev:2007qg} Abelev~B~I et al (STAR Collaboration) 2007 
{\it Phys. Rev.} C {\bf 75} 1054906

\bibitem{Teaney:review} Teaney~D~A 2010 
{\it Quark-Gluon Plasma 4\/} edited by R~Hwa and X-N~Wang  
(World Scientific, Singapore) 207 

\bibitem{Kolb:2003dz} Kolb~P~F and Heinz~U 2003 
{\it Quark-Gluon Plasma 3\/} edited by R~Hwa 
(World Scientific, Singapore) 634 

\bibitem{Huovinen:2006jp} Huovinen~P and Ruuskanen~P~V 2006 
{\it Annu. Rev. Nucl. Part. Sci.} {\bf 56} 163

\bibitem{Heinz:2009xj} Heinz~U 2010
{\it Relativistic Heavy Ion Physics\/} Landolt-Boernstein {\bf 23} 
edited by Stock~R (Springer, New York) 240

\bibitem{Romatschke:2009im} 	
 Romatschke~P 2010 {\it Int. J. Mod. Phys.} E {\bf 19} 1

\bibitem{Plumari_2012}
PlumariaS, Puglisib~A, Scardinab~F and Greco~V 2012
{\it Phys. Rev.} C {\bf 86} 054902 

\bibitem{Wiranata_2012} Wiranata~A and Prakash~M 2012
{\it Phys. Rev.} C {\bf 85} 054908

\bibitem{Andronic_2012} Andronic~A 2012
\textit{A thermal fit of ALICE hadron yields} ALICE-ANA-421  

\bibitem{Becattini_2012} Becattini~F et al 2012 
{\it Phys. Rev.} C {\bf 85} 044921

\bibitem{Stock_2013}
Stock~R et al 2013 {\it PoS CPOD2013} 011

\bibitem{Knudsen} 
Knudsen~M 1909 {\it Ann. Phys. (Leipzig)} {\bf 28} 75

\bibitem{Borg}
Borghini~N 2006 {\it Eur. Phys. J.} A {\bf 29} 27

\bibitem{Gombeaud} Gombeaud~C and Ollitrault~J-Y 2008
{\it Phys. Rev.} C {\bf 77} 054904

\bibitem{Hirano:2005xf}
Hirano~T, Heinz~U, Kharzeev~D, Lacey~R and Nara~Y 2006
{\it Phys. Lett.} B {\bf 636} 299

\bibitem{Hirano:2010jg} Hirano~T, Huovinen~P and Nara~Y 2011
{\it Phys. Rev.} C {\bf 83} 021902

\bibitem{Song:2010mg}
Song~H, Bass~S~A, Heinz~U, Hirano~T and Shen~C 2011 
{\it Phys. Rev. Lett.} {\bf 106} 192301

\bibitem{Nagle:2011uz} Nagle~J~L, Bearden~I~G and Zajc~W~A 2011
{\it New J. Phys.} {\bf 13} 075004

\bibitem{Shen:2011eg} Shen~C, Heinz~U, Huovinen~P and Song~H  2011
{\it Phys. Rev.} C {\bf 84} 044903

\bibitem{Qiu:2011iv} Qiu~Z and Heinz~U 2011
{\it Phys. Rev.} C {\bf 84} 024911

\bibitem{PhysRevC.76.041903} Drescher~H-J and Nara~Y 2007 
{\it Phys. Rev.} C {\bf 76} 041903

\bibitem{Miller}
Miller~M, Reygers~K, Sanders~S~J and Steinberg~P 2007
{\it Annu. Rev. Nucl. Part. Sci.} {\bf 57} 205


\bibitem{Kolb:2000sd} Kolb~P~F, Sollfrank~J and Heinz~U~W  2000 
{\it Phys. Rev.} C {\bf 62} 054909

\bibitem{Luzum:2009sb} Luzum~M and Romatschke~P 2009
{\it Phys. Rev. Lett.} {\bf 103} 262302

\bibitem{PHOBOSeccPART} Alver~B et al (PHOBOS Collaboration) 2008
{\it Phys. Rev.} C {\bf 77} 014906

\bibitem{PhysRevC.68.034903} Alt~C et al (NA49 Collaboration) 2003 
{\it Phys. Rev.} C {\bf 68} 034903

\bibitem{Aharony}
Aharony~O, Gubser~S~S, Maldacena~J~M, Ooguri~H and Oz~Y 2000
{\it Phys. Rep.} {\bf 323} 183

\bibitem{TSchaef_14} Sch{\"a}efer~T 2014 arXiv:1403.0653 [hep-ph]



\bibitem{Jia_1} Jia~J and Mohapatra~S 2013
{\it Phys. Rev.} C {\bf 88} 014907 

\bibitem{Jia_2} Huo~P, Jia~J and Mohapatra~S 2014
{\it Phys. Rev.} C {\bf 90} 024910

\bibitem{Jia_3} Jia~J 2014 
{\it J. Phys. G: Nucl. and Part. Phys.} {\bf 41} 124003 

\bibitem{Yan_14} Yan~Li, Ollitrault~J-Y and Poskanzer~A~M 2015
{\it Phys. Lett.} B {\bf 742} 290

\end{thebibliography}
\end{document}